\documentclass[aps, nofootinbib,  preprint, 12pt]{revtex4}

\usepackage{graphicx}
\usepackage{dcolumn}
\usepackage{bm}
\usepackage{amssymb,amsmath,amsfonts,amsbsy,epsfig}
\usepackage{color}

\def\fun#1#2{\lower3.6pt\vbox{\baselineskip0pt\lineskip.9pt
 \ialign{$\mathsurround=0pt#1\hfil##\hfil$\crcr#2\crcr\sim\crcr}}}

\def\lsim{\mathrel{\rlap{\lower4pt\hbox{\hskip1pt$\sim$}}
    \raise1pt\hbox{$<$}}}         
\def\gsim{\mathrel{\rlap{\lower4pt\hbox{\hskip1pt$\sim$}}
    \raise1pt\hbox{$>$}}}

\usepackage{cancel}

\def\beq{\begin{equation}}
\def\eeq{\end{equation}}                         
\def\bea{\begin{eqnarray}}
\def\eea{\end{eqnarray}} 
\def\to{\rightarrow}  
                           
\begin{document}

\preprint{IPMU-09-0104}
\title{Neutrino mass from a hidden world and its phenomenological implications}
\author{Seong Chan Park}
\author{Kai Wang}
\author{Tsutomu T. Yanagida}
\affiliation{Institute for the Physics and Mathematics of the Universe, University of Tokyo, Kashiwa, Chiba 277-8568, JAPAN}

\begin{abstract}

We propose a model of neutrino mass generation in 
extra dimension. 
Allowing a large lepton number violation on a distant brane spatially 
separated from the standard model brane, a
small neutrino mass is naturally generated due to an exponential
suppression of the messenger field in the 5D bulk. 
The model accommodates a large Yukawa coupling with the
singlet neutrino ($n_R$) which may change
the standard Higgs search and can simultaneously 
accommodate visible lepton number violation
at the electroweak scale, which leads to very interesting phenomenology 
at the CERN Large Hadron Collider.

\end{abstract}
\maketitle

\section{Introduction}
Enormous experimental evidence clearly indicates that neutrinos have tiny but non-zero
masses, and understanding the origin of neutrino masses is one 
of the most pressing problems in particle physics. 
The minimal Higgs boson model with additional right-handed neutrinos 
provides a simple solution to all the fermion masses including neutrino masses
by generating all of them through Yukawa interactions. However, the $10^{12}$ order 
hierarchy between dimensionless Yukawa coupling $y_t$ and neutrino Yukawa coupling $y_\nu$
suggests that neutrino masses may arise from an additional source besides the electroweak
symmetry breaking. Their electric neutrality  allows for the possibility of 
neutrinos being Majorana fermions. 
Within the Higgs boson model framework,
the seesaw mechanism \cite{GRSY, M} is an elegant proposal
accounting for the tiny neutrino masses. It is crucial that 
the small neutrino masses arise as a consequence of the grand unification 
at ultra high energy scales \cite{GRSY,so10}. The mechanism is based on the 
presence of singlet heavy Majorana neutrinos of mass $m_R$,
\beq
y \overline{l_L} n_R \tilde{H}+m_R \overline{n^c_R} n_R~,
\eeq
which is impossible
to probe directly at the collider experiments.
However, there have recently been several proposals to show how to test the origin
of neutrino mass directly at the coming CERN Large Hadron Collider (LHC) \cite{lhc}.
In the seesaw mechanism, the tiny neutrino masses arise as a consequence of
lepton number violation at the ultra high energy scale while if the new physics
responsible for neutrino mass is accessible at the LHC, additional tuning
may be needed. 

In the effective theory language, the Majorana
neutrino mass after integrating out new physics at a scale $\Lambda$ is due to the term
\beq
y^2 l_L l_L HH {S^n\over \Lambda^{n+1}}\label{llhh}~,
\eeq
where $S$ is a dimension one scale and $y$ is a dimensionless Yukawa coupling. Therefore, without tuning $y$ in Eq. (\ref{llhh}), there are only two approaches   
to generate a tiny Majorana neutrino mass. One approach is to impose
discrete (gauge) symmetries which generate a large value of  $n$ and tiny
neutrino masses arise as high dimensional operators \cite{discrete}.
Another approach is to have a small scale $S$. For instance,
in the ``inverse seesaw'' models \cite{inverse} or ``TeV triplet'' models \cite{typeii}, 
to obtain the correct neutrino mass, a keV order
scale needs to be introduced, and this small scale may be identified as
a soft breaking of Lepton symmetry \cite{'tHooft:1979bh}.
Within the framework of minimal type-I seesaw \cite{GRSY,M}, where $n=0$ in Eq.(\ref{llhh}),
if the heavy Majorana neutrinos are of the electroweak scale, the only way to generate the 
correct neutrino mass is to require the Yukawa coupling $y$ to be of $\mathcal{O}(10^{-6})$. 
Due to the tiny Yukawa coupling, 
the production of $n_R$ can only 
be enhanced through new gauge interactions such as with a $U(1)_{\rm B-L}$ gauge 
boson $Z_{\rm B-L}$ or $SU(2)_R$ gauge boson $W_R$ \cite{zbl}.

\begin{figure}[h]
\includegraphics[width=0.645\textwidth]{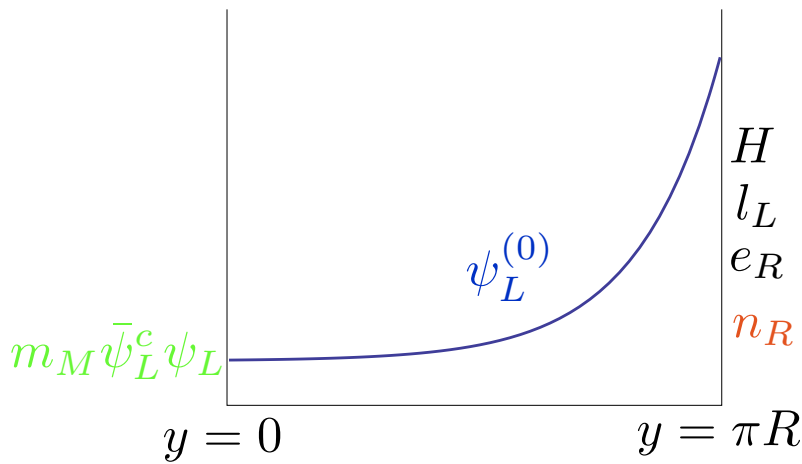}  
\caption{Five dimensional setup of the present model. All the standard model particles plus one singlet neutrino $n_R$ are localized on the visible brane located at $y=\pi R$. The lepton violating sector is on a distant brane located at $y=0$ and has only small overlap with the zero-mode of a messenger field ($\psi_L^{(0)}$), which induces a small lepton number violation transmuted to the visible sector and results a small neutrino mass.}
\label{Fig:extra dimension}
\end{figure}

In this paper, we propose a new model based on extra dimension \cite{extra_d}. Here, neutrino mass is generated by a cooperation with a right-handed singlet field $n_R$ on ``visible brane'' and a lepton number violation at a distant brane, ``hidden brane,'' spatially separated in the extra dimension communicating through a messenger field in the 5D bulk (See Fig. \ref{Fig:extra dimension}).
With a messenger field whose zero-mode wave function has an exponential
suppression at the hidden brane,
even if the lepton number violation on the hidden brane is as large as the electroweak scale, the resultant
neutrino mass remains naturally small. 
Differently from existing models in extra dimension \cite{models_extrad}
our model predicts two phenomenologically interesting features 
(i) a large Yukawa coupling with $n_R$, which can 
completely change Higgs phenomenology
(ii) a sizable lepton number violation through mixings with Kaluza-Klein excitation modes 
of the messenger field, which can be tested by future collider experiments such as
the CERN Large Hadron Collider (LHC). 

\section{Model}
The space-time in the model is a five dimensional spacetime with an orbifold
extra dimension $S^1/Z_2$ whose fixed points are located at $y=0$ and $y=\pi R$. 
Two branes are introduced at the end points: ``hidden brane" at $y=0$ and  ``visible brane" at
$y=\pi R$. All the standard model particles including leptons ($l_L=(\nu_L, e_L)^T, e_R$) and 
Higgs field ($H$) are localized on the visible brane with an additional singlet right-handed fermion ($n_R$), hence the particle spectrum is the same as the one in the conventional $SO(10)$ GUT model.  Lepton number is a good symmetry on the visible brane.  However a large lepton number violation is allowed on the hidden brane and it can communicate with the standard model sector through a messenger field,  $\Psi (x,y)=\psi_L+\psi_R$, in the 5D bulk. Here, $\psi_{L/R}=(1\pm\gamma_5)\Psi/2$. As the minimal spinor representation in 5D is vector-like so that a 5D bulk field has both of chiralities. 
Imposing lepton number $+1$ for the messenger field the violation of lepton
number is effectively parameterized by the localized  Majorana mass of
the messenger field ($m_M$).
Furthermore, the $Z_2$ transformation of the messenger field is defined as
$$Z_2: \Psi(x,y)\to \Psi(x,-y)=\gamma_5 \Psi(x,y)$$
so that $\psi_L$ has even parity satisfying Neuman boundary conditions at the end points ($y=0$ and $\pi R$) and conversely $\psi_R$ has odd parity satisfying Dirichlet boundary conditions thus vanishes at the end points.  Accordingly, only $\psi_L$ can have a non-vanishing Majorana mass term, $\overline{\psi_L^c}\psi_L \delta(y)$ at $y=0$. $\psi_R$, on the other hand, could have neither boundary localized Majorana mass term $\overline{\psi^c_R}\psi_R \delta(y)$ nor  couplings with the SM Higgs field  at $y=\pi R$.   One should also notice that only the left-chiral 
state ($\psi_L^{(0)}\sim e^{m_\Psi y}$) has a zero-mode. KK excitation modes consist of massive Dirac spinors ($\psi_L^{(n>0)},\psi_R^{(n>0)}$) having Kaluza-Klein masses $m_n^2 = m_\Psi^2+n^2/R^2$.

The action of the model is given as:
\begin{eqnarray}
S_5&& =\int d^4 x \int_0^{\pi R}dy  \,\,  \overline{\Psi} i \Gamma^M D_M \Psi - m_\Psi\overline{\Psi}\Psi  \nonumber
\\
&&+\delta(y-\pi R) \left({\cal L}_{\rm SM} + y \overline{l_L}\tilde{H} n_R + m_D \overline{\psi}_L n_R  \right) \nonumber
\\
&& + \delta(y) \left( m_M \overline{\psi_L^c} \psi_L + H.C. \right).
\end{eqnarray}
Notice that lepton number is  violated only on the hidden brane and the amount of violation is effectively parametrized by a Majorana mass parameter $m_M$.

After the Kaluza-Klein decomposition 
$\psi_{L/R} (x,y) = \sum_n \psi^{(n)}_{L/R}(x) f_{L/R}^{(n)}(y)$ 
and integrating out the fifth dimension, we get the  4D effective Lagrangian with a tower of Kaluza-Klein states as 
\begin{eqnarray}
&&{\cal L}_{\rm eff} 
\ni \delta \bar{\nu}_L n_R + \Delta_0 \bar{\psi}_L^{(0)}n_R +\sum_{n,m \geq 0} \epsilon_{nm}\bar{\psi^c}_L^{(n)} \psi_L^{(m)} \nonumber \\
&&+\sum_{n>0}\Delta_n \bar{\psi}_L^{(n)}n_R + m_n (\bar{\psi}_L^{(n)}\psi_R^{(n)}+\bar{\psi}_R^{(n)}\psi_L^{(n)})~.
\label{lag}
\end{eqnarray}
Here, we have introduced  convenient parameters $\delta = y \langle H \rangle,  \epsilon_{nm}=\epsilon_{mn}= m_M f_L^{(n)}(0) f_L^{(m)}(0)$ , $\Delta_n = m_D f_L^{(n)}(\pi R)$ where $f_L^{(n)}$ is the wave function of the $n$-th Kaluza-Klein mode $\psi_L^{(n)}.$ One should notice that  $\epsilon_{00}$ is much smaller than $\epsilon_{nm}$ with a non-zero 
$n$ and/or $m$:
$$\epsilon_{0 0} \ll \epsilon_{0 n>0}\ll \epsilon_{n>0, m>0}$$
due to the large exponential suppression $f_L^{(0)}(0)^2 \sim  e^{-2\pi R m_\Psi}$ as we have assumed $m_\Psi>1/R$. This exponential suppression is the very essence ensuring a small neutrino mass at the end.

In the basis of $(\nu_L, \psi_L^{(0)}, n_R^c, \psi_L^{(1)}\psi_R^{(1)c}, \psi_L^{(2)}, \psi_R^{(2)c}\cdots)$ the mass matirx is given as

\begin{eqnarray}
{\cal M}=\left(
  \begin{array}{ c c  c c c c  c c }
     0 & 0 & \delta & 0 & 0 &0 &0 &\cdots \\
     0 & \epsilon_{00} & \Delta_0 &\epsilon_{01} & 0 &\epsilon_{02}&0&\cdots \\
     \delta & \Delta_0& 0 & \Delta_1 &0& \Delta_2&0&\cdots \\
          0 & \epsilon_{10}& \Delta_1 & \epsilon_{11} & m_1 &\epsilon_{12}&0&\cdots \\
     0 & 0& 0 & m_1 & 0 &0&0&\cdots \\
     0&\epsilon_{20}&\Delta_2&\epsilon_{21}&0& \epsilon_{22}&m_2 &\cdots\\
     0&0&0&0&0&m_2&0&\cdots \\
     \vdots & \vdots & \vdots & \vdots & \vdots &\vdots&\vdots & \ddots
  \end{array} \right).
  \label{5D matrix}
\end{eqnarray}
Interestingly, the determinant of this mass matrix can be easily calculated thanks to lots of zeroes in the matrix coming from lepton number conservation on the visible brane, the Dirichlet boundary condition for $\psi_R^{(n)}$ and the orthogonality of KK states, as
\begin{eqnarray}
\det{\cal M} = \epsilon_{00}\delta^2 \Pi_i m_i^2.
\end{eqnarray}
Obviously, if $\delta$ or $\epsilon_{00}$ vanishes this determinant vanishes thus the neutrino remains exactly massless. Since $\epsilon_{00}$ is exponentially small in our model the smallness of neutrino mass is achieved.

\section{Phenomenology}
The key feature of the model is that an exponential suppression of
the zero-mode wave function of the messenger field 
leads to the small Majorana mass $\epsilon_0$ of $\psi^{0}_L$
while the small $\epsilon_0$ naturally 
leads to the small neutrino masses even when 
the Dirac mass parameter $m_D$ or Yukawa coupling constant $y$ ($\delta= y \langle H \rangle$) 
of $n_R$ are sizable. 
Consequently, the model predictions
can then be directly tested or constrained by experiments. 

Assuming all the mass parameters are in the electroweak scale, $1/R$ controls the lepton number violation in the heavy neutrino states and may lead to totally different collider signature based on different choices of the parameter region.
We first consider the case when the compactification scale is similar to the weak scale ($1/R\sim M_W$). In this case KK states directly come into play in low energy phenomenology. 
To illustrate the feature, we choose a set of model parameters as an example as 
$$(1/R, m_\Psi, m_M, m_D, \delta)=(200, 800, 1000, 500, 20)\,{\rm GeV},$$
then the mass spectrum is given by
\begin{eqnarray}
&&(m_\nu, m_-, m_+, m_4, m_5, \cdots) \nonumber \\&&=(2.4\times 10^{-10}, 181, 193, 360, 859, \cdots)\, {\rm GeV}.\nonumber
\end{eqnarray}
a sizable mass splitting between the lightest two heavy states arises. 
The splitting of the Majorana masses will lead to visible Lepton number violation in the heavy neutrino states.
The direct search of such heavy neutrinos at the LHC are widely studied in \cite{lhc} in completely model independent
phenomenological approach. If the KK states are not decoupled,  the current model becomes an explicit model realization for this signature. 

If the compactification scale is much higher than the electroweak scale ($1/R \gg M_W$), only three light modes, $(\nu_L, \psi^{(0)}_L, n^c_R)$, are directly relevant in low energy phenomenology.  The neutrino mass matrix is reduced to a simple $3\times 3$ matrix 
\begin{eqnarray}
{\cal M}\to {\cal M}_{\rm eff}\simeq \left(
  \begin{array}{ c c  c}
     0 & 0 & \delta\\
     0 & \epsilon_0 & \Delta_0 \\
     \delta & \Delta_0 & 0
  \end{array} \right)~.
  \label{4D matrix}
\end{eqnarray}
In this limit, the model essentially reduces to conventional ``inverse seesaw'' models in 4D    \cite{inverse} where the smallness of $\epsilon_0$ can be argued as a soft breaking of Lepton symmetry \cite{'tHooft:1979bh}. In this model, the small Majorana mass ($\epsilon_0$) is guaranteed by  the higher dimensional nature of the model in a natural way. 
 At ${\cal O}(\epsilon_0)$, 
the eigenstates are one light state ($\nu$) and two nearly degenerate heavy
states ($N_\pm$):
\begin{eqnarray}
\nu &\approx& \frac{\Delta_0}{\sqrt{\Delta_0^2+\delta^2}}\nu_L -\frac{\delta}{\sqrt{\Delta_0^2+\delta^2}} \psi^{(0)}_L, \\
N_\pm &\approx& \frac{1}{\sqrt{m_\pm^2 +\Delta_0^2+\delta^2}}(m_\pm n_R^c +\delta \nu_L +\Delta_0 \psi^{(0)}_L)
\label{Eq:mass eigenstate}
\end{eqnarray}
with corresponding mass eigenvalues 
\begin{eqnarray}
m_\nu \approx \epsilon_0 \frac{\delta^2}{\delta^2+\Delta_0^2}, \,\,
m_\pm \approx \sqrt{\delta^2+\Delta_0^2}\pm {\cal O}(\epsilon_0).
\label{Eq:masses}
\end{eqnarray}

The Lepton number violation effects are suppressed as $\epsilon_0/\sqrt{\delta^2+\Delta_0^2}$.
The heavy neutrinos are
dominant by their Dirac component and direct search of heavy Dirac neutrino 
has been studied in \cite{dirac} in the framework of Type-III seesaw.


We want to emphasize that the Yukawa 
coupling $y$ can be of  $\mathcal{O}(1)$ in the model.  
Therefore, one may expect immediate implication in Higgs phenomenology 
since Higgs can  decay into neutrino states as
\beq
H\to \overline{\nu} N_\pm,
\eeq
provided these decays are allowed kinematically.
Then, $N_\pm$ states can decay into $\nu b\bar{b}$ through a virtual Higgs.

Before going to details of Higgs study, we will discuss the physical states and
mixing constraints first. The above mixings contribute to $\mu$-decay by the effective muon decay constants as 
\begin{eqnarray}
G_\mu\simeq G_F (1-\frac{1}{2} \frac{\delta_\mu^2}{\Delta_0^2})(1-\frac{1}{2} \frac{\delta_e^2}{\Delta_0^2}).
\end{eqnarray}
As a consequence, all the observables which depend on the Fermi constant will be affected by the mixing angles or $\delta_e/\Delta_0$ and $\delta_\mu/\Delta_0$. Also if $\Delta_0<M_Z$, the invisible decay rate of $Z^0$ can be reduced by $(1-\frac{1}{6}\sum_l \delta_l^2/\Delta_0^2)$ with respect to the standard model one. Taking $\mu-e$ universality, CKM unitarity and the invisible decay rate of $Z^0$ one can get a results at $90\%$ C.L. \cite{bound1,bound2,bound3} as
\begin{eqnarray}
\frac{\delta_{e,\mu,\tau}^2}{\Delta_0^2}  < (6.5, 5.9, 18)\times 10^{-3}.
\end{eqnarray}

If Higgs is light as $m_H \lesssim 140$ GeV,  it dominantly decays into $b\bar{b}$ due to
the limited allowed phase-space. However the bottom Yukawa coupling is only $y_b = m_b/\langle H\rangle \simeq 1.7\times 10^{-2}$.
 If heavy neutrino Yukawa coupling $y$ is sufficient large
and the Higgs has a decay phase space, the Higgs decays into neutrino, $H\to \nu N$, 
may siginificantly changes the standard Higgs search procedure. 
The partial decay width is 
\beq
\Gamma(H\to \nu^i_L \bar{N^j}) = \frac{|{y}_{ij}V_{ij}|^2}{8\pi}m_H\left(1-{m^2_N\over m^2_H}\right)^2,
\eeq
where $m_H$, $m_N$ are masses of Higgs and heavy neutrino respectively and $V_{ij}$ is the mixing
fraction of the neutrino state $n^i_R$ in the physical states $N^j$. 
To simplify the discussion, we focus on the ``inverse seesaw'' limit where $N_+$ and $N_-$ are nearly
degenerate and 
\beq
n_R\simeq (N_++N_-)/\sqrt{2}~~;m_{N_+}\simeq m_{N_-}~.
\eeq 
Then, the total
width for Higgs decaying into heavy neutrino states is given by
\begin{eqnarray}
&&\sum \Gamma(H\to\nu N_\pm) \\
&&= {1\over 2 |V_{R+}|^2}\Gamma(H\to \nu N_+)+{1\over 2 |V_{R-}|^2}\Gamma(H\to \nu N_-). \nonumber 
\end{eqnarray}
As discussed previously, 
there are strong bounds on the heavy neutrino mixing both from 
direct search experiments and precision test of electroweak interactions. For 
$m_N < $100 GeV, LEP experiments L3 and DELPHI  provided bounds 
as $\delta_l/\Delta_0< 10^{-2}$ \cite{bound:LEP} and hence 
$y \sim m_N/\langle H\rangle \times \delta_l/\Delta_0 < 4\times 10^{-3}$.
The phase space suppression in $H\to \nu N$ is mostly larger than in $H\to b\bar{b}$.
We don't expect the standard model Higgs decay branching fraction to have visible change in this case.
If $m_N$ is large enough to escape from LEP bounds, the precision test on electroweak interaction put less stringent bound on $\delta_\tau^2/\Delta_0^2<0.018$. $y$ can then
be as large as $\mathcal{O}(10^{-2})$. Even though the $y$ can be a few times
bigger than $y_b$, the heavy neutrino channel has a much larger decay phase space
suppression. When the mass difference between $m_H$ and $m_N$ are sufficiently
large of $30\sim 40$ GeV, however, the new channel will significantly reduce the BR($H\to b\bar{b}$) and BR($H\to \gamma\gamma$). 

\begin{figure}[h]
\includegraphics[width=0.648\textwidth]{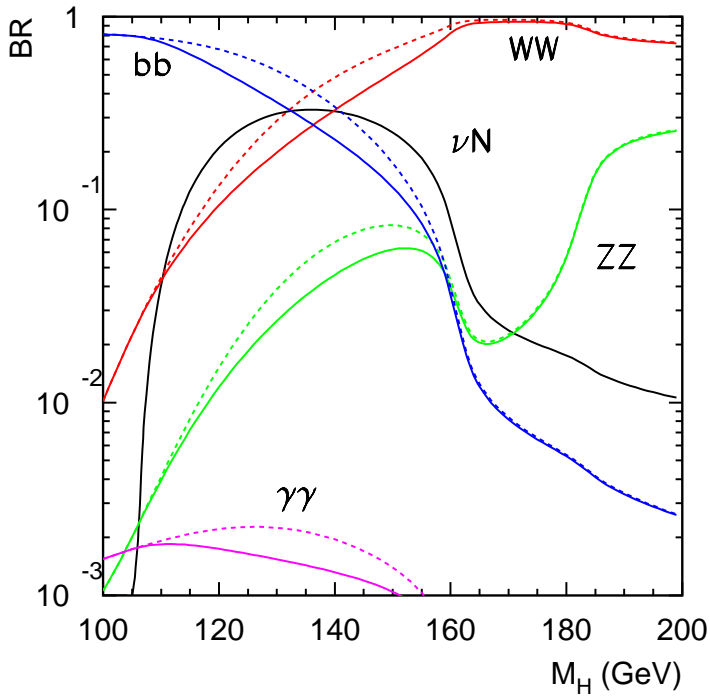}
\caption{Modified Decay BR of the SM Higgs for 
$y=6\times 10^{-2}$, $m_N$=105 GeV}
\label{br}
\end{figure}

To illustrate the qualitative feature, we assume there is only one generation heavy neutrino state that have large
Yukawa coupling and plot the Higgs decay BR figure for $m_N=105$ GeV, $y=6\times 10^{-2}$ in Fig. \ref{br}.
The dash-lines and solid-lines correspond to 
the original Higgs decay BR without $H\to \nu N$ and the modified BR of
Higgs with $H\to \nu N$ decay, respectively. In principle, we will need three
generations of heavy neutrino states and this can significantly increase the 
partial width of Higgs decaying to heavy neutrinos.

The $H\to \nu N$ channel will be very challenging to be identified since
there is always a missing neutrino. Then, it is impossible to fully reconstruct the Higgs.
Since the Higgs decay may not be dominantly into
heavy neutrinos, the conventional searching channels are
still available. However, notice that some decay BRs 
are significantly changed as in Fig. ~\ref{br}.
Due to the new decaying channel, for instance Br($H\to \gamma\gamma$)
can be reduced by a factor more than of 50\% while
as argued in \cite{Low:2009di}.

\section{Conclusion and Remarks}

In this paper we suggest a new model of neutrino which only involves TeV scale masses. 
The lightness of neutrinos is guaranteed by an exponential suppression in the
zero-mode wave function of a messenger field in 5D bulk which mediates lepton number violation taking place on a distant brane separated from the brane where all the standard model leptons are confined. 
Depending on parameter choice of $1/R$, the model can accommodate both the electroweak scale Majorana neutrino \cite{lhc} models and the ``inverse seesaw'' type model \cite{inverse}. 
The model naturally accommodates a large Yukawa coupling in $\bar{l} n_R H$ and may lead to interesting phenomenology
in the Higgs search. 

In the end, we want to add another remark regarding on the sizable Lepton number violation effects at the electroweak scale.
Even if one allows a large $m_R \overline{n^c_R}n_R$ term on the visible brane, the smallness of a light neutrino is guaranteed by the tiny Majorana mass $\epsilon_0$.
Heavy states ($N_\pm$), on the other hand, have a sizable mass splitting 
\begin{eqnarray}
m_\pm && \approx \sqrt{\delta^2+\Delta_0^2+ \frac{m_R^2}{4}}\pm \frac{m_R}{2}\pm {\cal O}(\epsilon_0),
\end{eqnarray}
and the lepton number violation effects are now
of order $m_R/\sqrt{\delta^2+\Delta^2+ \frac{m_R^2}{4}}$. Thus, if $m_R$
is as large as the electroweak scale, one will expect significant lepton number violation
effects at the LHC. 

\section*{Acknowledgement}
The work supported by the World Premier International Research Center Initiative (WPI Initiative), MEXT, Japan. 


\end{document}